\pdfoutput=1
\documentclass[12pt]{iopart}
\usepackage{graphicx}
\usepackage{dcolumn}
\usepackage{caption}

\newcommand{ \be }{\begin{eqnarray}}
\newcommand{ \ee }{\end{eqnarray}}
\newcommand{ \ben }{\begin{enumerate}}
\newcommand{ \een }{\end{enumerate}}
\newcommand{ \la }{\langle}
\newcommand{ \ra }{\rangle}

\newcommand{ \eps }{\varepsilon}

\usepackage{color}

\begin{document}
\title{Anisotropic flow:  Achievements, Difficulties, Expectations}

\author{Sergei A. Voloshin}

\address{Department of Physics and Astronomy\\
Wayne State University, Michigan 48201 USA }

\ead{voloshin@wayne.edu}

\begin{abstract}
Anisotropic flow measurements play a crucial role in understanding
the physics and bulk properties of the system created in heavy ion
collisions. In this talk I briefly review the most important results obtained
so far, recent developments in the analysis techniques and the
interpretation of the results, 
and what should we expect next, both at RHIC and LHC.
I also discuss event anisotropies sensitive to the strong parity
violation effects.
\end{abstract}


Analysis of the event anisotropies (anisotropic flow) 
in multiparticle production in non-central
nuclear collisions appeared to be one of the most informative paths in
understanding the physics and characterizing the properties of the
dense and hot strongly interacting medium. It has been observed a
continuous increase in the value of in-plane elliptic flow
($v_2>0$)~\cite{Barrette:1996rs} 
from the top AGS energies to RHIC.
At RHIC, strong elliptic flow~\cite{Ackermann:2000tr}
comparable in strength to the predictions of ideal hydrodynamics, 
and the hadronization via quark coalescence following from constituent
quark number scaling of differential 
flow~\cite{Voloshin:2002wa,Molnar:2003ff}, 
together with the jet quenching, are the key
ingredients in the picture of sQGP (strongly coupled Quark Gluon
Plasma).

The field is rapidly developing and evolving. From the ``discovery
phase'' of the first years of RHIC operations it has been transforming into 
detailed quantitative
description of the sQGP phase and subsequent hadronization.
The plots of $v_2/\eps$ vs particle density~\cite{Voloshin:2002wa,
Alt:2003ab,Voloshin:2007af}
 that have been extensively used in 
the assessment of the level of thermalization reached in the 
relativistic nuclear collisions
and applicability of ideal hydrodynamics comes under scrutiny: 
are the measurements of elliptic flow precise
enough, are the anisotropies what we think they are and how much are they
modified by fluctuation processes? When comparing to
hydrodynamical 
calculations, are the proper initial conditions used in the calculations?  
Could it be that we have missed some important physics building the models?
In the last couple years significant progress has
been reached answering each and every of these questions. 
The role of viscosity, flow fluctuations, initial conditions
(eccentricity and initial flow (velocity) field) are a few questions
to report in this review. I also discuss preliminary measurements of
azimuthal
 ``out-of-plane'' anisotropies that are related to one of
the fundamental questions of the strong parity violation.  
Finally, I briefly overview future measurements at RHIC and LHC.
Unfortunately, the space does not allow any detailed discussion of many 
other very important developments,
such as flow of $\phi$-mesons~\cite{Afanasiev:2007tv,Abelev:2007rw} 
that is important to understand the
relative flow development in the partonic and hadronic phases, 
flow of the deuterons~\cite{Afanasiev:2007tv} 
and $K^*$-mesons as tests of coalescence, 
heavy flavor flow, etc. The KE scaling of elliptic flow observed by
PHENIX Collaboration~\cite{Adare:2006ti}
is still not fully understood/appreciated.
There is no doubt that these measurements will enrich our understanding
of the dense QCD medium even further.


Taking into account the significance of the $v_2/\eps$ plot in 
establishing the sQGP picture, 
I spend most of my time discussing recent developments related to
this plot. There have been  a number of important findings:
(a) Along with several indirect indications that even in central Au+Au
collisions the thermal equilibrium is not complete, it was found that 
even very small, compared to the conjectured low limit of $\eta/s$
values, the viscous effects lead to a significant reduction 
in the predicted elliptic
flow compared to the ideal hydro case. This would lead to a
contradiction with experimental measurements if other effects,
responsible for an increase of elliptic flow (compared to the
``standard'' hydrodynamical calculation) would not be identified. 
Several of such effects have been reported.
(b) First, it was noticed that ideal hydro calculations, if tuned to
describe spectra, yield larger elliptic flow than thought
previously.
(c) It was shown, that in some models, e.g. CGC, the initial eccentricity   
can take significantly larger values than in optical Glauber model
that is usually used in hydro calculations. 
The larger eccentricities inevitably lead to larger elliptic flow. 
(d) Flow fluctuations, the nature of which is much better understood
in the last years, lead to an increase of {\em apparent} flow. 
(e) Finally,
it was noticed that the gradients in the initial velocity
field also lead to the increase in final values of elliptic flow.
Any of the above mentioned effects can be quite significant, leading each to
20-30\% or even larger change in values of $v_2$. 
The final ``assembly'' of these
effects into one reliable model is still under way.

An attempt of model independent 
analysis of $v_2/\eps$ dependence on particle density based on
parametrization in terms of {\em Knudsen number} has been developed 
in~\cite{Bhalerao:2005mm,Drescher:2007cd}. 
Using an expression
$v_2/\eps =(v_2/\eps)_{hydro} (1+K/K_0)^{-1}$ (where the parameter
$K_0\approx 0.7$ is independently estimated from comparison
 to model calculations)
to fit the data, see Fig.~1, the authors conclude that at 
RHIC we might be still up to 30\% below the ideal ``hydro limit'' even for
the most central collisions. 
Their estimate of the viscosity yields values of
$\eta/s=0.11-0.19$ depending on the CGC or Glauber initial conditions.
Similar fits to the STAR data performed by
R.~Snellings~\cite{Raimond:private}  and
collaborators lead to similar conclusions.

\begin{figure}[t]
\begin{minipage}[t]{0.48\textwidth}
  \includegraphics[width=0.95\textwidth]{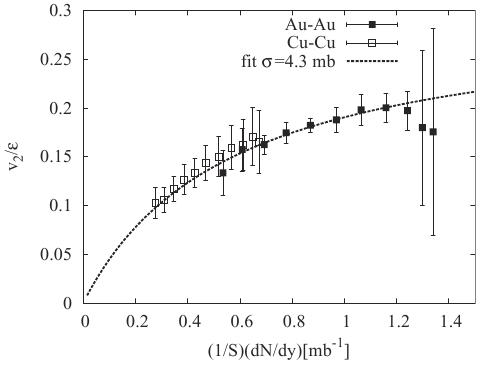}
\label{fig1}
\caption{Fit to $v_2/\eps$ in terms of Knudsen number~\cite{Drescher:2007cd}.}
\end{minipage}
\hspace{0.03\textwidth}
\begin{minipage}[t]{0.48\textwidth}
  \includegraphics[width=0.95\textwidth]{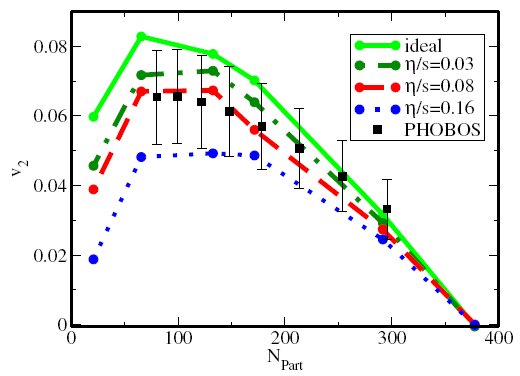}
\caption{Viscous hydro calculations~\cite{Romatschke:2007mq}
compared to data.}
\label{fig:romat}
\end{minipage}
\end{figure}

\begin{figure}[t]
\begin{minipage}[t]{0.48\textwidth}
  \includegraphics[width=0.95\textwidth]{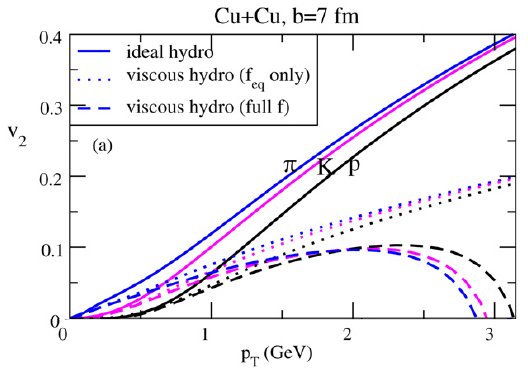}
\caption{
$v_2(p_t)$ from ideal and  viscous ($\eta/s=1/(4\pi)$) hydrodynamics~
\cite{Song:2007ux} }.
\label{fig:song}
\end{minipage}
\hspace{0.03\textwidth}
\begin{minipage}[t]{0.48\textwidth}
  \centerline{\includegraphics[width=0.9\textwidth]{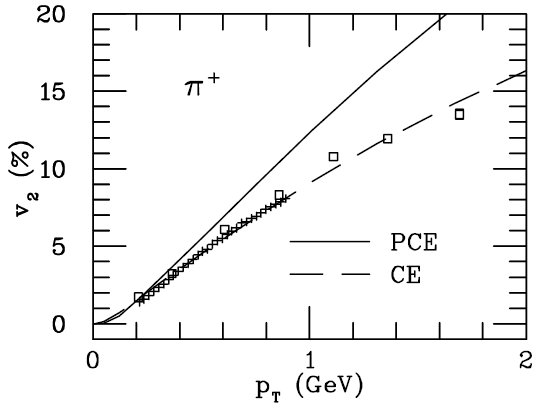}}
\caption{Pion $v_2(p_t)$ in two scenarios~\cite{Huovinen:2007xh}. 
Solid line indicates the
results using parameters best fit to spectra.}
\label{fig:pasi}
\end{minipage}
\end{figure}

The magnitude of {\em the viscous effects} could be judged already from the
early calculations~\cite{Teaney:2001av}
 where the hydro dynamical evolution at some intermediate stage was joined to
the transport model to simulate late (viscous) evolution of the
system. Later, viscosity was attempted to be introduced directly into
hydrodynamic calculations~\cite{Teaney:2003kp}. 
Recently there have been performed a few ``full''
calculations~\cite{Romatschke:2007mq,Song:2007fn,Song:2007ux}
of the hydrodynamical expansion with viscous terms
explicitly included into equations. 
Note that the latter (namely the
form of these terms) is not that everybody agrees
on~\cite{Song:2007fn}, though the difference in the results due to use
of somewhat different equations are likely small.  
At the same time everybody agrees on the significance of the viscous 
effects even for the ``minimal'' values of viscosity ($\eta/s=1/(4\pi)$). 
The results presented in Figs.~\ref{fig:romat} and~\ref{fig:song} 
show that even
the minimal viscosity lead up to $\sim$25-30\% reduction in flow
values in Au+Au collisions and probably more than 50\% in Cu+Cu.

Note that viscosity
coefficients calculated in pQCD are usually much larger than 
would be allowed by the data. 
In this sense, noteworthy are the recent calculations~\cite{Xu:2007jv}, 
which emphasize the importance of taking into account  
$2 \leftrightarrow 3$ processes.  
With these effects included,
the viscosity coefficient appeared to be about an order magnitude smaller
compared to previous calculations
and fall into the ``allowable'' by the data range.

One can wonder how is it possible that with viscous
effects to be that strong that 
the ideal hydrodynamical calculations results are not much higher
than the experimental measurements? 
Indeed, there have been identified several effects which possibly lead
to a significant increase in predicted flow. Taken together with viscous
effects they may restore the agreement with experiment.

Firstly, in~\cite{Huovinen:2007xh} it was explicitly demonstrated 
that the ideal hydro calculations can not be ``tuned''
to describe both spectra and elliptic flow. 
If one tunes the
model to describe spectra, the elliptic flow values appeared too
large, for about 20-30\%, compared to the data, see Fig.~\ref{fig:pasi}. 
What leaves even more space for viscous effects, is the
observation~\cite{Hirano:2005xf,Drescher:2006pi} that
the initial eccentricity calculated in the CGC model yields 
values up to 50\% larger compared to the ``standard'' optical Glauber
model. The effect has been further studied
in~\cite{Drescher:2007ax}; in~\cite{Lappi:2006xc} it was
discussed how eccentricity depends on details of CGC model, which can be taken
as possibility to investigate CGC model by measuring flow.

The role of flow fluctuations and non-flow effects is another
long standing problem that received a lot of attention and  
significant progress has been made in the recent
couple years. In particular, the role of fluctuations
in  the initial system geometry~\cite{Miller:2003kd}  defined by nuclear 
{\em participants}~\cite{Manly:2005zy} 
(interacting nucleons or quarks) has been greatly
clarified~\cite{Voloshin:2006gz,Bhalerao:2006tp,Voloshin:2007pc,Alver:2008zz,Broniowski:2007ft}. 
The following picture emerges:
Anisotropic flow is defined as the correlations to the reaction plane
spanned by the impact parameter and the beam axis. 
At fixed impact parameter, the geometry of the {\em participant zone}
fluctuates, both, in terms of the value of the
eccentricity as well as the orientation of the major
axes. 
Then the anisotropy develops along the plane spanned by the minor
axis of the participant zone and the beam direction, 
the so called {\em participant plane}. As the true
reaction plane is not known and the event plane is estimated from
the particle azimuthal distribution ``defined'' by the participant plane, 
the apparent (participant plane) flow appears to be  always
bigger (and always ``in-plane'', $v_{2,PP}>0$) compared to the ``true''
flow as projected onto the reaction plane (see Fig.~\ref{fig:planesQ}).  

It was noticed
in~\cite{Voloshin:2007pc} that in collisions of heavy nuclei the
fluctuations in the eccentricity 
$(\eps_x,\eps_y)=(\la (\sigma_y^2-\sigma_x^2)/(\sigma_y^2+\sigma_x^2)\ra,
\la (2\sigma_{xy}^2/(\sigma_y^2+\sigma_x^2)\ra)$ 
can be well described by two-dimensional Gaussian. 
What is not trivial is that for such Gaussian
fluctuations the higher cumulant flow ($v\{n\}, n\ge 4$) is not only 
insensitive to non-flow but also to eccentricity fluctuations. 
All of higher cumulants are
exactly equal to the ``true'' flow, namely as given by projection onto
the reaction plane. At the same time, the apparent (participant plane)
flow become unmeasurable in a sense that flow fluctuations could not
be separated from non-flow contributions by means of correlation
measurements.       

An important conclusion from that
study was  that in most cases (except, probably, in mid-peripheral
and peripheral Cu+Cu collisions, when the Gaussian approximation
breaks~\cite{Alver:2008zz}, and $\eps_{part}\{4\}$ does not agree with 
$\eps_{std}$~\cite{Voloshin:2006gz})   the measurements of higher cumulant
flow values provide the elliptic flow
relative to the true reaction plane. 
Similarly, the same value
is given by several other methods, such as Lee-Yang Zeroes, Bessel
Transform, and fit to q-distributions. 
This greatly simplifies the comparison, e.g. of hydrodynamical calculations
to the data, as it says that in such calculation one should not
worry how to take into account the fluctuations in the initial
eccentricity (which is a non-trivial task) but just compare to the
``right'' measurement, e.g. $v_2\{4\}$.  
This understanding also allowed to appreciate some earlier calculations
with uRQMD model~\cite{Zhu:2005qa,Zhu:2006fb}. There, it was shown
that using higher cumulants and/or LYZ method one indeed can measure
the elliptic flow very well, but it was not at all clear why there
have been observed no traces of the effects of flow fluctuations, 
which were expected in this model.  

Unfortunately this progress in understanding the nature of fluctuations
does not help in resolving the problem of {\em measuring} 
flow fluctuations (in the participant plane) and non-flow. 
Strictly speaking, to make any estimates of those  one is required to
make assumptions.  
Most often to suppress non-flow contribution the azimuthal correlations
between particles with  large rapidity separation are used.     
The problem with this method is that there is no reliable estimates of
how well it suppresses non flow and also how much the flow
fluctuations (in this case correlations) change after imposing such a cut. 

\begin{figure}[t]
\begin{minipage}[t]{0.48\textwidth}
  \includegraphics[width=0.95\textwidth]{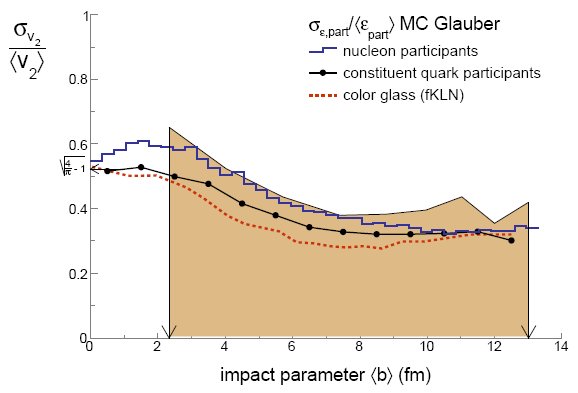}
\caption{Elliptic flow fluctuations as estimated by STAR.}
\label{fig:flucStar}
\end{minipage}
\hspace{0.03\textwidth}
\begin{minipage}[t]{0.48\textwidth}
  \includegraphics[width=0.95\textwidth]{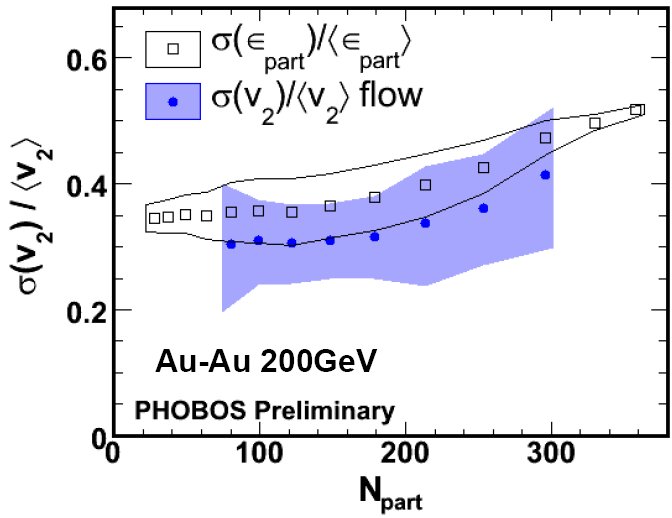}
\caption{Elliptic flow fluctuations as estimated by PHOBOS
Collaboration.}
\label{fig:flucPhobos}
\end{minipage}
\end{figure}

\begin{figure}[t]
\begin{minipage}[t]{0.48\textwidth}
  \includegraphics[width=0.95\textwidth]{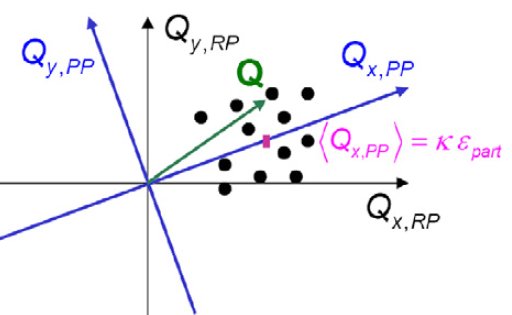}
\caption{Flow vector distribution at 
fixed $(\eps_v,\eps_y)$~\cite{Voloshin:2007pc}}
\label{fig:planesQ}
\end{minipage}
\hspace{0.03\textwidth}
\begin{minipage}[t]{0.48\textwidth}
  \includegraphics[width=0.95\textwidth]{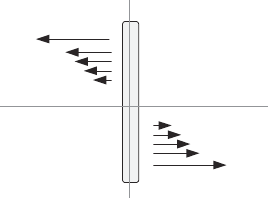}
\caption{Initial velocity profile in non-central nuclear 
collisions~\cite{Becattini:2007sr}}
\label{fig:becat}
\end{minipage}
\end{figure}

At this conference the PHOBOS~\cite{PHOBOSv2fluc} and the 
STAR~\cite{STARv2fluc} collaborations presented their
revised (compared to QM'06) results on flow fluctuations.
These results are in good qualitative and quantitative agreement,
see Figs.~\ref{fig:flucStar},~\ref{fig:flucPhobos}. 
In~\cite{STARv2fluc} a conservative approach is taken and only upper
limits on fluctuations are reported. The PHOBOS Collaboration 
uses estimates of non-flow effects from correlations
with large rapidity separations and report more restrictive range for
fluctuations.  
Both agree that the current
measurements exhaust the (nucleon) eccentricity values obtained 
in MC Glauber model and in
this sense somewhat favor models which predict smaller relative
fluctuations, such as the CGC model or MC Glauber taking into account
constituent quark substructure. 

Another important and interesting direction that is just started 
to be explored is the role of the 
non-zero initial flow velocity profile. 
The first obvious candidate here is to take into account the initial
velocity  gradient along the impact parameter, Fig.~\ref{fig:becat}. 
As shown in~\cite{Becattini:2007sr} such a gradient directly contributes to the
in-plane expansion rate (see Eq.~22 in~\cite{Becattini:2007sr}). 
I would draw attention to the fact that those effects naturally
also lead to {\em directed} flow (see the same Eq.~22), which too briefly
addressed in~\cite{Troshin:2007cp}. 
It will be very interesting to compare the calculations in such a model to 
a very precise data from STAR~\cite{Wang:2007kz} 
on directed flow obtained with the
reaction plane determined by neutrons in the Zero Degree Calorimeter.
The relation to other models~\cite{Csernai:2006yk,Snellings:1999bt}
predicting non-trivial dependence of directed flow on rapidity would
be also very interesting. 
Speculating on this subject one would also notice that viscous effects
must also play an important role in such a scenario. 

\begin{figure}[t]
\begin{minipage}[t]{0.48\textwidth}
  \includegraphics[width=0.95\textwidth]{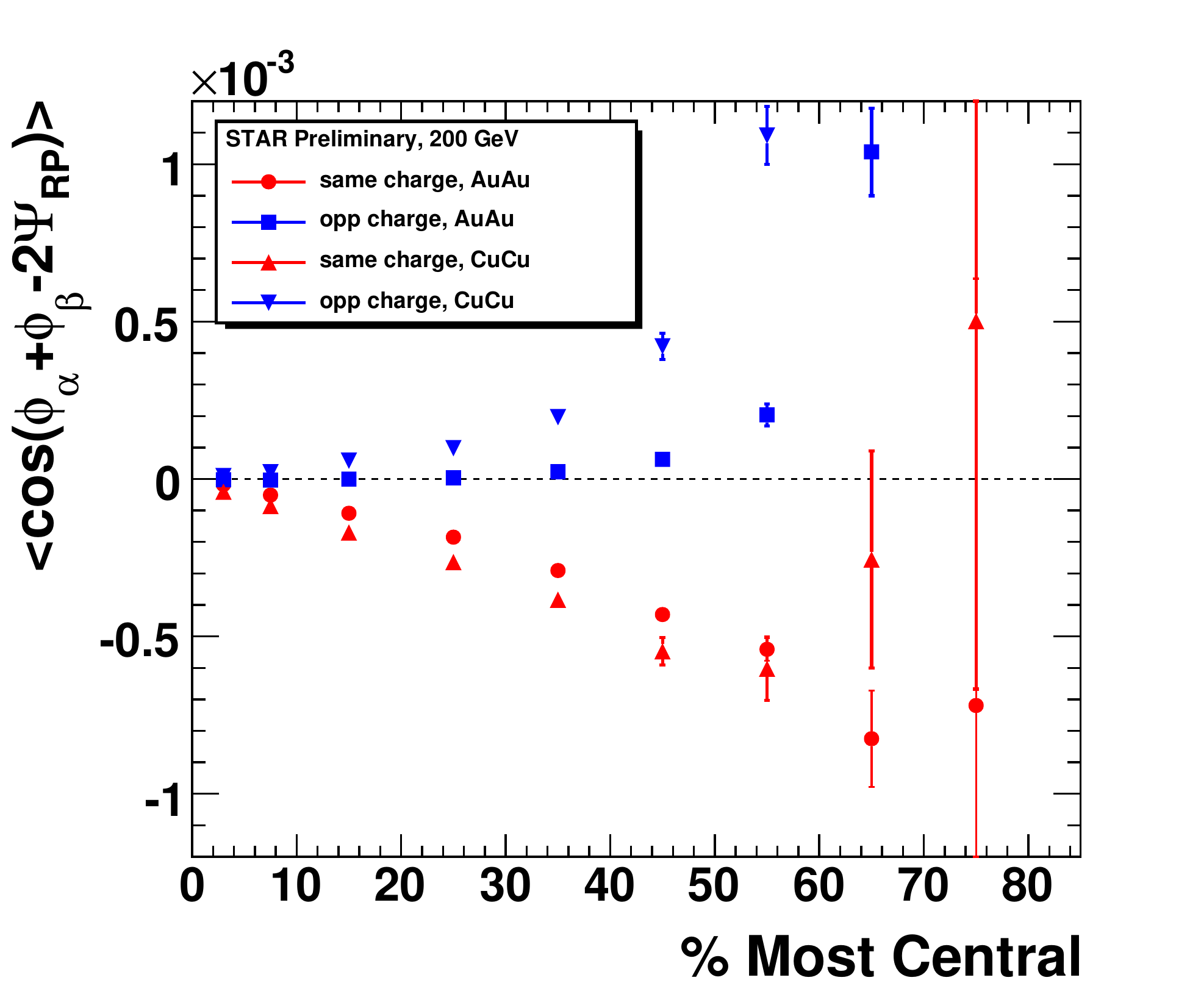}
\caption{Azimuthal anisotropy correlator sensitive to strong 
$\cal P-$violation effects.}
\label{fig:parity}
\end{minipage}
\hspace{0.03\textwidth}
\begin{minipage}[t]{0.48\textwidth}
  \includegraphics[width=0.95\textwidth]{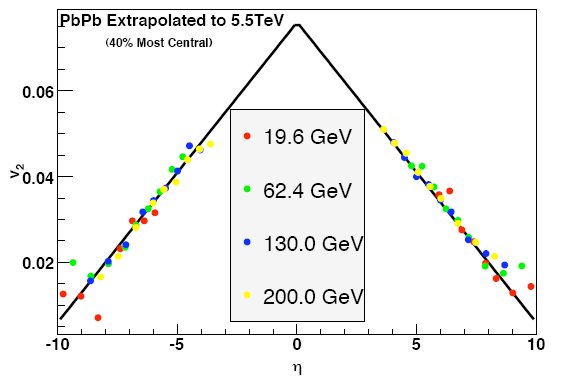}
\caption{Extrapolation of $v_2$ values to LHC
energies~\cite{Busza:2007ke}.} 
\label{fig:busza}
\end{minipage}
\end{figure}

It was shown in~\cite{Kharzeev:2004ey} 
 that the effect of the strong parity violation that lead
to non equal number of right and left fermions (quarks) 
in the presence of strong magnetic fields of colliding nuclei
would result in charge separation (preferential emission of same
charge particles)  in the direction perpendicular to
the reaction plane. 
Such anisotropy, which very much resembles
``out-of-plane directed flow'' can be addressed with the
help of three-particle correlations~\cite{Voloshin:2004vk}
by measuring $\la \cos(\phi_\alpha+\phi_\beta-2\Psi_{RP}) \ra$,
where $\phi_{\alpha,\beta}$ are azimuthal angles of two (same or
 opposite) charged particles, and $\Psi_{RP}$ is the reaction plane angle. 
The estimates~\cite{Kharzeev:2007jp} (see also talk by H.~Warringa at
 this conference)
indicate that the effect is
 strong enough to be observed in heavy ion collisions. 
At this conference the STAR Collaboration reported
the preliminary results~\cite{Voloshin:qm2008}, 
see Fig.~\ref{fig:parity}, 
that qualitatively agree with
estimates presented in~\cite{Kharzeev:2004ey,Kharzeev:2007jp}. 
Note, that the correlator 
$\la \cos(\phi_\alpha+\phi_\beta-2\Psi_{RP}) \ra$  is $\cal P$-even and
contain contributions from other effects not related to parity
violation. 
A careful analysis of such a contribution is obviously
needed before any strong conclusion can be drawn from these measurements.

Coming years promise many new interesting data from low energy RHIC run
and, of course, from LHC. The main interest in the low energy RHIC scan, 
anisotropic flow is no exception, is the search for the
QCD critical point. The scan would cover the energy region from top
AGS energies, over the CERN SPS, and higher. In terms of
anisotropic flow two major observables to watch would be a possible
``wiggle'' in $v_2/\eps$ dependence on particle density~\cite{Voloshin:1999gs}
 and
``collapse'' of directed flow~\cite{Stocker:2007pd}.
 RHIC also has plans to extend its reach
in terms of energy density using uranium beams. From the first
estimates and ideas of using uranium beam we now have real detailed
simulations~\cite{Nepali:2007an} of such collisions 
with developed methods for a selection of
desired geometry of the collision. 

The predictions for the LHC are rather uncertain, though most agree
that the elliptic flow will continue to increase~\cite{Abreu:2007kv},
partially due relatively smaller contribution of viscous effects.  
Simple extrapolations~\cite{Busza:2007ke,Borghini:2007ub} of
the collision energy dependence of $v_2$ look rather ''reliable'. 
Note that there exist calculations predicting {\em decrease} of the
elliptic flow~\cite{Krieg:2007sx}. 
Another important observation is an increase in mass dependence
(splitting) of $v_2(p_t)$ due to a strong increase of radial flow.

In summary, we have had very exciting years of anisotropic flow study,
which greatly enriched our understanding of ultra-relativistic nuclear
collisions and multiparticle production in general. We are looking
forward for new physics with LHC and RHIC.

\begin{figure}[t]
\begin{minipage}[t]{0.48\textwidth}
  \includegraphics[width=0.95\textwidth]{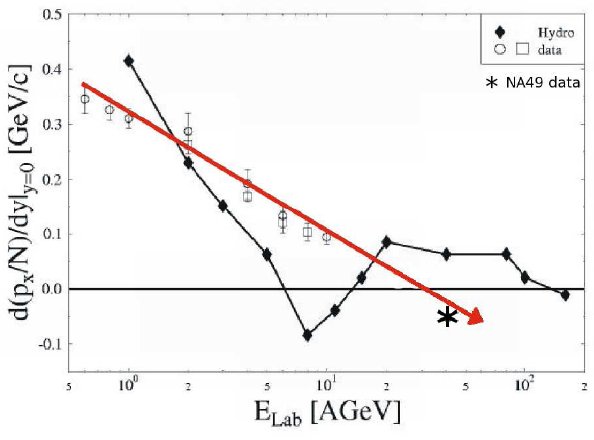}
\caption{``Collapse'' of directed flow as discussed in~\cite{Stocker:2007pd}.}
\end{minipage}
\hspace{0.03\textwidth}
\begin{minipage}[t]{0.48\textwidth}
\end{minipage}
\end{figure}

\vspace*{1mm}
I thank A. Poskanzer, R. Snellings, and A. Tang for numerous fruitful
discussions.

\vspace*{3mm}

\end{document}